\documentclass{PoS}
\usepackage{epsfig}
\PoS{PoS(BDMH2004)063}

\title{IMF variations and their implications for Supernovae numbers}

\ShortTitle{IMF variations}

\author{\speaker{Carsten Weidner}\\
        Sternwarte der Universität Bonn, Germany\\
        E-mail: \email{cweidner@astro.uni-bonn.de}}

\author{Pavel Kroupa\\
        Sternwarte der Universität Bonn, Germany\\
        E-mail: \email{pavel@astro.uni-bonn.de}}

\abstract{The stellar initial mass function (IMF) integrated over
an entire galaxy is an integral over all separate star-formation events.
Since most stars form in star clusters with different masses the integrated
IMF becomes an integral of the (universal or invariant) canonical
stellar IMF over the star-cluster mass function. This integrated IMF
is steeper (contains fewer massive stars per G-type star) than the
canonical stellar IMF. Furthermore, observations indicate a relation
between the star-formation rate of a galaxy and the most luminous
stellar cluster in it. This empirical relation can be transformed into
one between the star-formation rate of a galaxy and a maximum cluster
mass. The assumption that this cluster mass marks the upper end of a
young-cluster mass function leads to a connection of the
star-formation rate and the slope of integrated IMF for massive stars.
This integrated IMF varies with the star-formation history of a
galaxy. Notably, large variations of the integrated IMF are evident
for dwarf galaxies. One important result is that the number of type II
supernovae per star is
supressed relative to that expected for a canonical IMF, and that dwarf
galaxies have a supressed number of supernovae per star relative to massive
galaxies. For dwarf galaxies the number of supernovae per star also varies
substantially depending on the galaxy assembly history.}

\FullConference{Baryons in Dark Matter Halos\\
		 5-9 October 2004\\
		 Novigrad, Croatia}

\begin{document}

\section{The stellar IMF}
Star formation takes place mostly in embedded clusters,
each cluster containing a dozen to many million stars. Within these
clusters stars appear to form following a 
universal initial mass function (IMF) with a Salpeter power-law slope
or index
($\alpha=2.35$, \cite{Sa55,Kr02}) for stars more massive than $1\,M_\odot$,
  $\xi(m) \propto m^{-\alpha}$, where $\xi dm$ is the number of
  stars in the mass interval $m, m+dm$. This has
been found to be the case for a wide range of different conditions in
the Milky Way (MW), the Large and Small Magellanic clouds
(respectively LMC, SMC) and other galaxies.

\section{The (embedded) cluster IMF}
Several studies show that star clusters also seem to be
distributed according to a power-law embedded
cluster mass function (ECMF), $\xi_{\rm ecl} = k_{\rm ecl} M_{\rm ecl}^{-\beta}$,
where $dN_{\rm ecl} = \xi_{\rm ecl}(M_{\rm ecl})~dM_{\rm
  ecl}$ is the number of embedded clusters in the mass interval
$M_{\rm ecl}$, $M_{\rm ecl} + dM_{\rm ecl}$ and $M_{\rm ecl}$ is
the mass in stars. In the solar neighbourhood \cite{LaLa03} find a slope 
$\beta = 2$ between 50 and 1000 $M_{\odot}$, while in the SMC and LMC
\cite{HuEl03} find $\beta \sim 2 - 2.4$ and \cite{ZhFa99}
find $\beta \sim 2 - 2.4$ for $10^4 \le M_{\rm ecl}/M_\odot \le 10^6$ in
the Antennae galaxies. We therefore assume a single-slope power-law
ECMF with $\beta = 2.35$ between 5 $M\odot$ and $10^{6}\,M_{\odot}$.

\section{The star-formation-rate--maximal-cluster-mass relation}
\begin{figure}[h!]
\centering
\epsfig{file=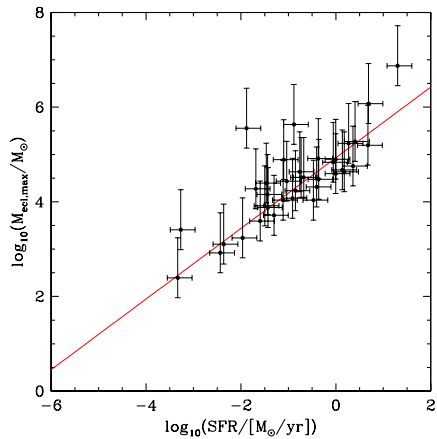,width=6cm}
\vspace*{-1.5cm}
\caption{The solid line shows the
  star-formation-rate--maximal-cluster-mass relation from
  \cite{WeKrLa04} while the dots with the error bars are the observations.}
\label{fig1}
\end{figure}

In \cite{WeKrLa04} we derived a relation (shown in Fig.~\ref{fig1})
between the maximal cluster 
mass in a galaxy and the current star formation rate (SFR) of the galaxy,
\[
\log_{10}({M_{\rm ecl,max}}) = \log_{10}({k_{\rm ML}}) + (0.75 \cdot
\log_{10}{\rm SFR}) + 6.77,
\]
where $k_{\rm ML}$ is the mass-to-light ratio, typically 0.0144 for
young ($<$ 6 Myr) clusters. This eq.~connects the IMF via the ECMF
with the properties of a galaxy. 
\vspace*{-1cm}

\section{The integrated galaxial initial mass function (IGIMF)}
\begin{figure}[h!]
\centering
\epsfig{file=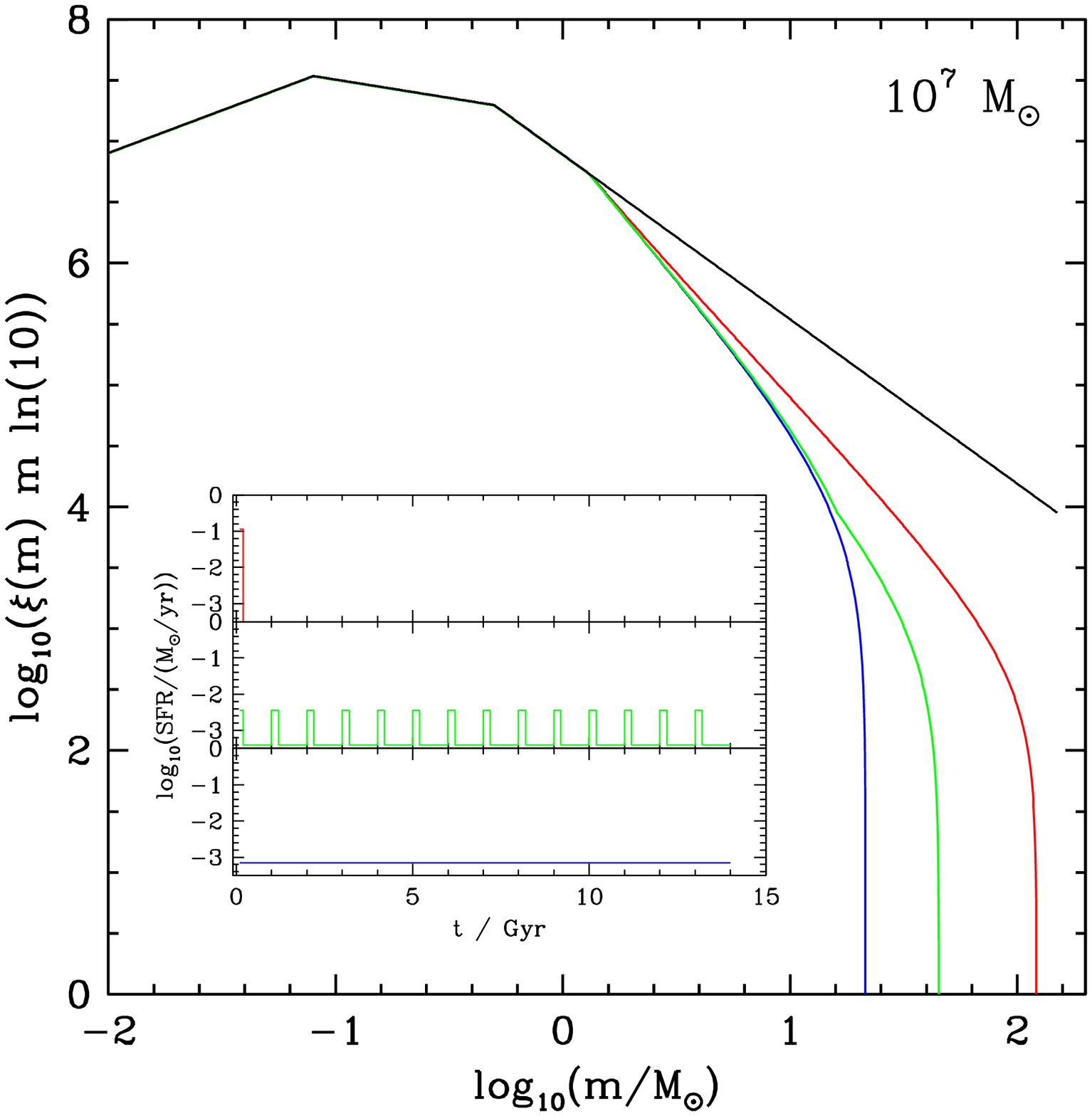,width=6cm}
\epsfig{file=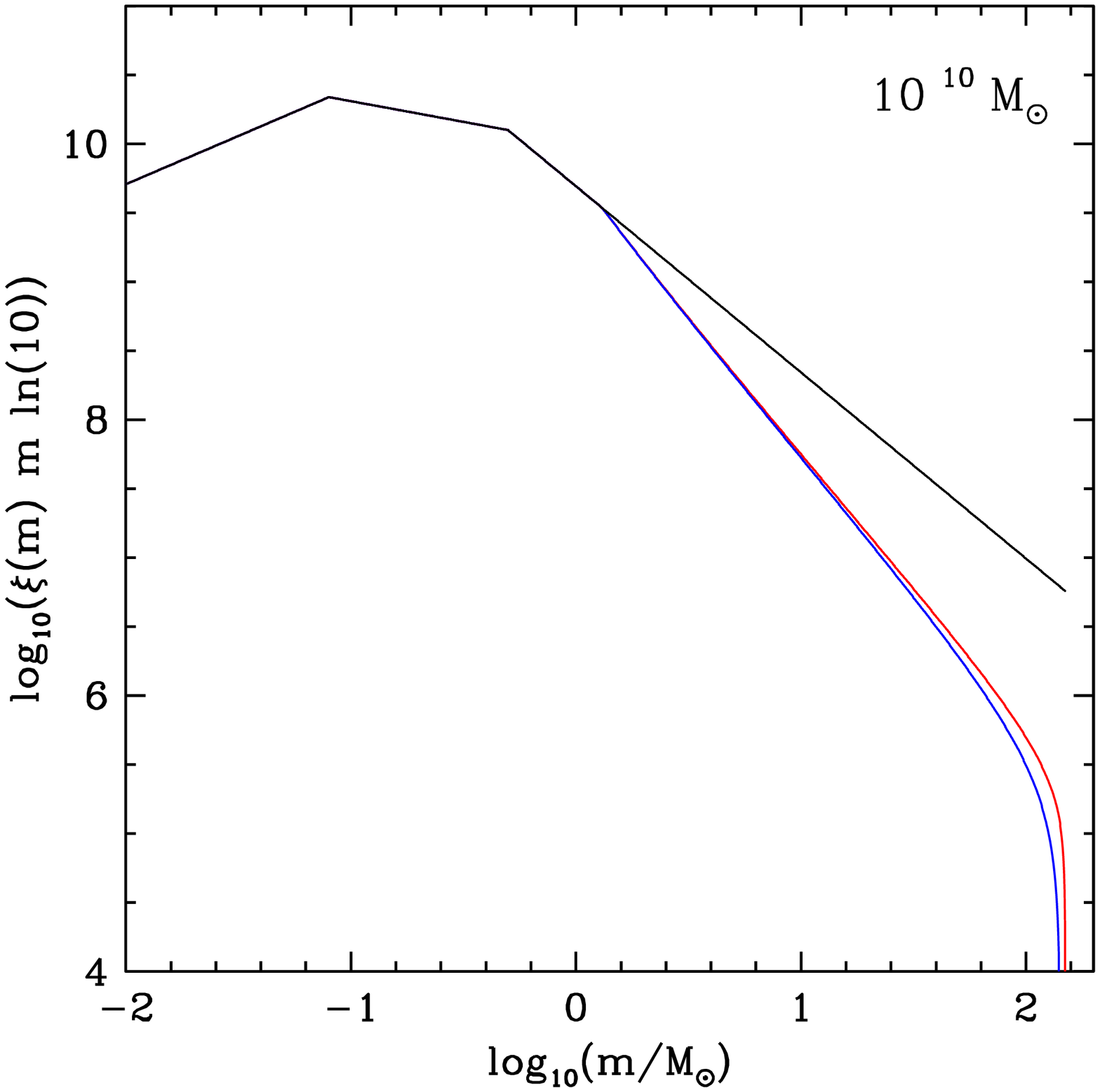,width=6cm}
\vspace*{-1.5cm}
\caption{Left panel: Integrated field mass functions
  for a galaxy with a stellar mass of $10^{7} M_{\odot}$ with different
  star-formation histories (single initial-burst of 100 Myr, periodic
  SF of 100 My every 900 Myr and constant SFR over 14 Gyr,
  respectively, from top to bottom) shown in the small box. The IMF
  slope above  
  1 $M_{\odot}$ is $\alpha_3 = 2.35$, while the ECMF slope
  $\beta$ is 2.35. Right panel: The same for a galaxy with
  $10^{10}\,M_{\odot}$ stellar mass. Both taken from \cite{WeKr04b}.}
\label{fig2}
\end{figure}

As in our model all stars are born in clusters following a universal IMF
but also clusters are formed from a universal ECMF, the mass function
for all stars born in all clusters, which we call the integrated galaxial
initial mass function (IGIMF), becomes the following integral (\cite{VB82}),
\[
\xi_{\rm IGIMF}(m) = \int_{M_{\rm ecl,min}}^{M_{\rm ecl,max}(SFR)} \xi(m\le
                      m_{\rm max})\, \xi_{\rm ecl}(M_{\rm ecl})\,dM_{\rm ecl}.
\]
As seen in Fig.~\ref{fig2} the resulting IGIMFs are always steeper
than the input canonical stellar IMF.

\begin{figure}[h!]
\centering
\epsfig{file=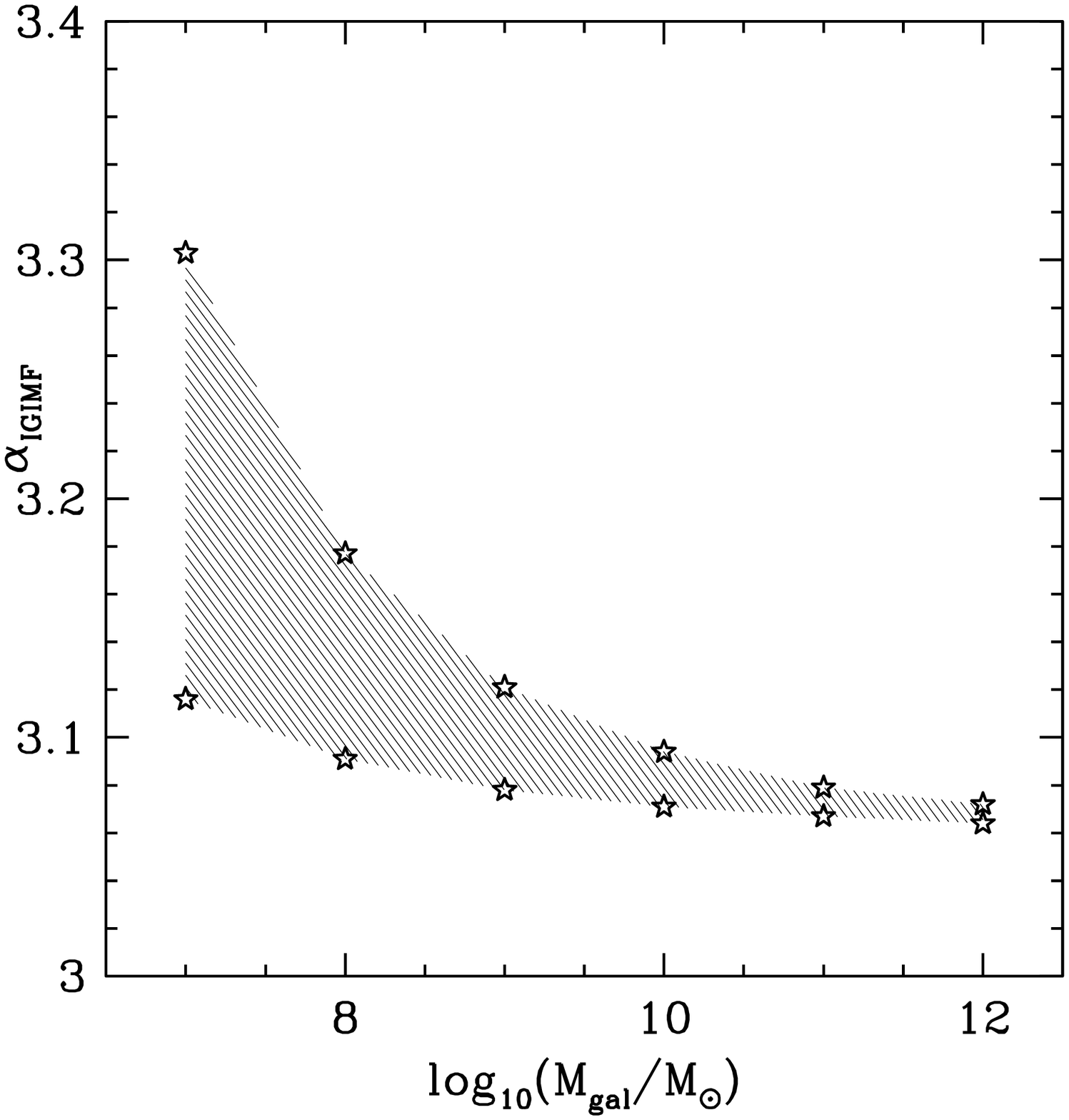,width=6cm}
\epsfig{file=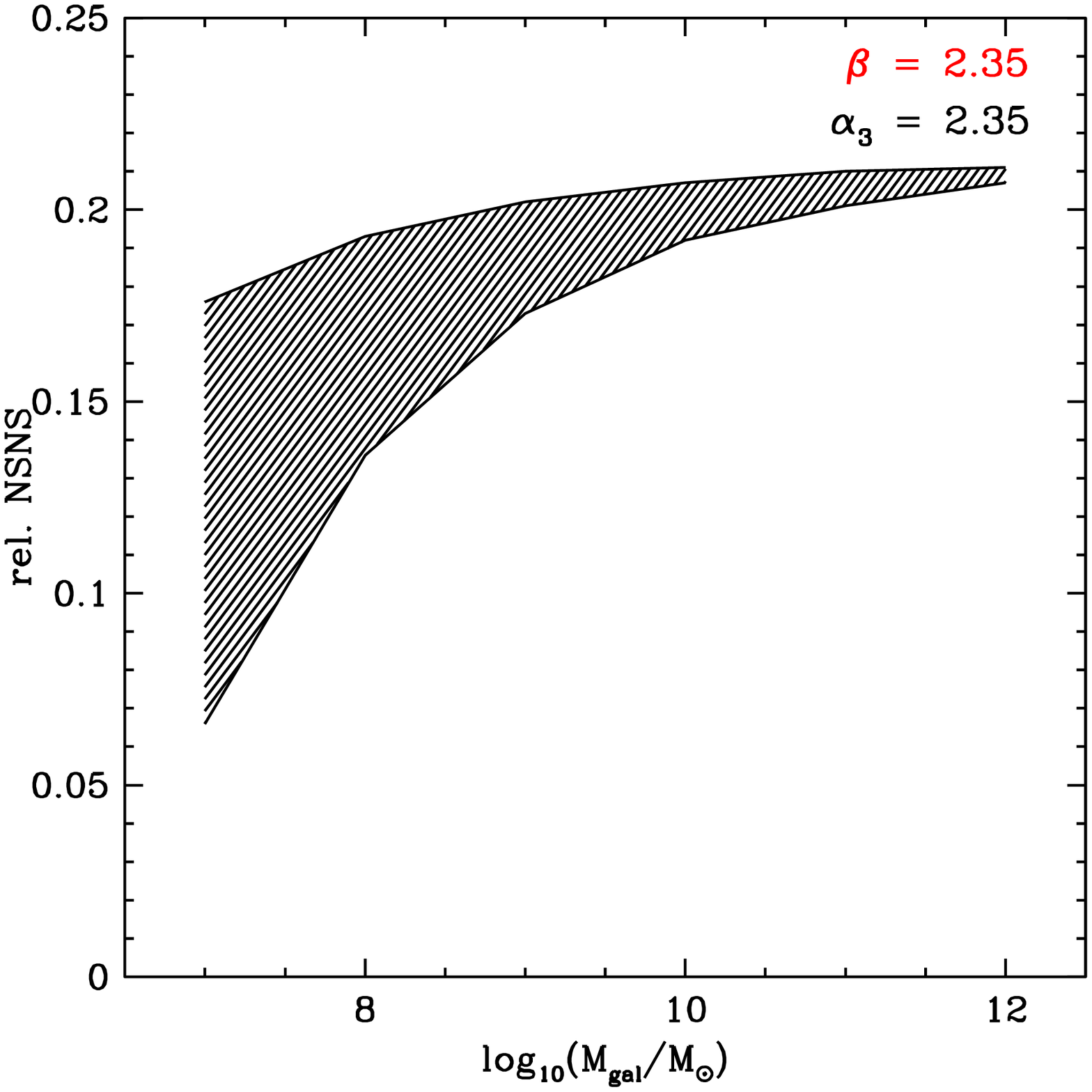,width=6cm}
\vspace*{-1cm}
\caption{Left panel: IGIMF slopes above $\sim 1.3\,M_{\odot}$ in
  dependence of the stellar galaxy mass for 
  $\alpha_{3} = 2.35$ and $\beta$ = 2.35. The lower limit of the
  shaded region results from single initial-burst models while the
  upper limit is from continuous models. The IGIMF thus becomes
  steeper with reducing stellar mass, thus resulting in systematic
  differences in the 
  chemical evolution of different galaxy types. Also in the low-mass
  regime a larger scatter in chemical properties is to be
  expected. Right panel: The number of SNII per star relative to the
  same number for a constant canonical IMF as a function of galaxy
  mass for $\alpha_{3}$ = 2.35 and $\beta$ = 2.35. The upper limit of
  the shaded region results from single burst models while the lower
  limit is deduced from continuous star-formation models. While for
  galaxies with large stellar masses the value is about 20\% of the
  corresponding canonical IMF value it drops significantilly for galaxies
  with lower stellar mass. Both panels are from \cite{WeKr04b}.} 
\label{fig3}
\end{figure}

\section{Results \& Conclusions}
\begin{itemize}
\item The IGIMFs must be steeper (see left panel of Fig.~\ref{fig3})
  than the stellar IMF and vary with galaxy type (see Fig.~\ref{fig2}).
\item Chemical enrichment histories, the number of SNII per star and
  mass-to-light ratios calculated with an invariant Salpeter IMF
  cannot be correct for any galaxy.
\item The number of supernovae per star (see right panel of
  Fig.~\ref{fig3}) is possibly significantly lower over 
  cosmological times than for an invariant canonical IMF.
\item Irrespective of how old a galaxy is it will always appear less
  chemically evolved than a more massive equally-old galaxy as a
  result of the steeper IGIMF.
\item The scatter in chemical properties must increase with decreasing
  galaxy mass.
\end{itemize}

\end{document}